\documentclass[prb,aps,twocolumn,eqsecnum]{revtex4}
\usepackage{mathtext}
\usepackage[T2A]{fontenc}
\usepackage[cp1251]{inputenc}

\usepackage{epsf}
\usepackage{psfrag}
\usepackage{graphicx}
\usepackage{amssymb,amsmath}

\begin{document}

\title{One dimensional light localization with classical scatterers; an advanced undergraduate laboratory experiment
}

\author{K.J. Kemp, S. Barker, J. Guthrie, B. Hagood, and M.D. Havey}%

\affiliation{\small Department of Physics, Old Dominion University,
Norfolk, VA 23529}

\email{mhavey@odu.edu}

\date{\today}

\sloppy

\begin{abstract}
The phenomenon of electronic wave localization through disorder was introduced by Anderson in 1958 in the context of electron transport in solids.  It remains an important area of fundamental and applied research.  Localization of all wave phenomena, including light, is thought to exist in a restricted one dimensional geometry.  We present here a series of experiments  which illustrate, using a simple experimental arrangement and approach, localization of light in a quasi one dimensional physical system.  In the experiments, reflected and transmitted light from a stack of glass slides of varying thickness reveals an Ohm's Law type behavior for small thicknesses, and evolution to exponential decay of the transmitted power for thicker slide stacks.  Light absorption is negligible in our realization of the experiment.  For larger stacks of slides, weak departure from a one dimensional behavior is also observed.  The experiment and analysis of the results, then showing many of the essential features of wave localization, is relatively straightforward, economical, and suitable for laboratory experiments at an undergraduate level.

\end{abstract}

\maketitle

\section{Introduction}
In a paper that has stimulated more than half a century of research, electron wave localization by disorder was introduced by Anderson \cite{Anderson1} in 1958.  In wave localization of all types, interferences that survive configuration averaging modify the transport properties of the wave. \cite{Sheng1,Akkerman1} In the ultimate case, transport can cease, leading to  fully spatially localized excitations.  In the case of electron transport through a solid medium, which is nominally a conductor, the material may transition to an insulator with increasing disorder.

Wave localization can be thought of as an emergent phenomenon.  That is, neither the fact of localization itself, or its macroscopic properties appear naturally as an evolution of ensemble properties transitioning from a collection of microscopic single elements to a macroscopic scale.  As a commentary on reductionism in the physical sciences, this idea is developed in the fascinating article by P.A. Anderson, $\emph{More is Different}$. \cite{Anderson2}

Localization of a number of different wave types has been observed, including electronic waves in solids, acoustic waves, \cite{Hefei} ultracold atomic waves \cite{Delande,Billy,Roati,Kondov} and microwaves. \cite{Chabanov}   Localization of nearly monochromatic light by a three dimensional disordered sample of classical scatterers has also been reported. \cite{Wiersma,Storzer} Comprehensive overviews of some of this research may be found in Ref. 2, 3, 6, 13 - 15.
Spatial localization of waves is sometimes termed strong localization.  There is a related interesting suite of phenomena that occurs under less stringent conditions, but which depend on similar interference effects as strong localization; these are collectively termed weak localization phenomena.  One well known example is the coherent backscattering effect, which was predicted 1984 by Kuga, $\emph{et al.}$ \cite{Kuga} and was demonstrated the next year by M. Van Albada, $\emph{et al.}$ \cite{Albada}.  Coherent backscattering (CBS) of light from a colloidal suspension has more recently been studied as part of an undergraduate level laboratory, \cite{Corey} where it provides a unique and accessible introduction to multiple coherent light scattering.  As was pointed out by Holcomb \cite{Holcomb} coherent backscattering of other waves, particularly electrons, also occurs, and makes a nice parallel with the case of CBS with light.  Weak localization of electrons may be studied at a similar level in thin silver films.\cite{Beyer}

Experimental studies of localization in two spatial dimensions can be generally challenging in execution and subtle in interpretation. \cite{Sheng1,Akkerman1,Delande}  In three spatial dimensions, there is additionally a critical amount of disorder required to invoke a localization transition. \cite{Sheng1,Akkerman1,Delande} For electron localization this can correspond to a conductor-insulator transition.  However, in one and quasi one dimension there are many instances of localization.  For example, it is well known that reflection from, for example, a roll of overhead projector film, a pile of microscope slides, or a stack of viewgraphs exhibits unusually strong reflection of visible light, \cite{Hecht} this being reminiscent of light localization in reflection.  For a roll of projector film, one can also readily observe strongly suppressed transmission at normal incidence to the roll, and yet ready transmission through the sides of the roll; the resulting circular ring of light is quite striking.  A more or less straightforward and quantitative demonstration of localization of light in a quasi one dimensional system can be made using a similar approach to that taken by Berry and Klein. \cite{Berry}  Although that study was concentrated mainly on the theory of light wave transport in a disordered medium in one dimension, an experimental confirmation of one of the main theoretical results was made. This was the exponential loss of transmitted intensity as the length of the scattering medium was increased. However, other observables such as the intensity of the backwards scattered light, the unitarity of the process as a whole, and direct comparison to the optical equivalent to Ohm's Law were not investigated.

With this background, the main purpose of this paper is to describe a set of experiments on quasi one dimensional light localization suitable for upper division physics majors.  The experiments may be combined into either a single experiment in a laboratory course, or laid out as a senior thesis project for one or a small team of undergraduates (as we have done).  The experiments are economical and are not technically complex, and yet provide convincing evidence of light localization in one dimension.  The experiments, along with the associated application of analysis to the results, open a door for upper level physics or optical engineering students into an important and very active area of contemporary research.

In the following sections we first describe in some detail our experimental set up. This is followed by experimental results showing clear evidence of light localization in quasi one dimension.  These results include the exponential decrease of transmitted power as the length of a sample is increased, the corresponding increase in the backwards scattering direction, and the departure of the transmitted light power from the optical equivalent of Ohm's Law.  Comparison of the decay exponent is made to the closed form results from Berry, $\emph{et al.}$ \cite{Berry}  We follow this with some comments on other aspects of these experiments and their analysis that may be studied by science undergraduates.

\section{Project Description}

\subsection{Experimental Approach}
\begin{figure}[th]
\begin{center}
{$\scalebox{0.9}{\includegraphics*{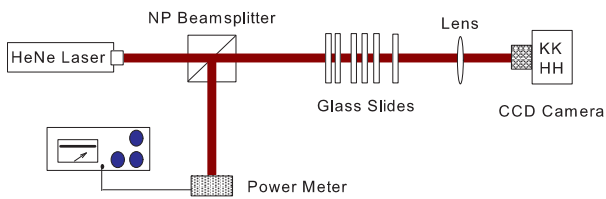}}
$ }
\caption{A schematic diagram of the experimental arrangement. CCD stands for a charge coupled device camera, while NP beamsplittter refers to a non polarization sensitive beam splitter. HeNe laser is a helium-neon laser.  }
\end{center}
\par
\label{fig1}
\end{figure}

A schematic diagram of the experimental apparatus is shown in Figure 1.   There the vertically polarized beam from a helium-neon (HeNe) laser is directed towards a stack of N glass slides.  The laser beam, of wavelength $\lambda$ = 628 $nm$, has a nearly Gaussian transverse intensity profile with an associated Gaussian radius of about $r_0$ = 0.5 $mm$. The intensity of the laser beam is about 0.5 $mW$.  The transverse profile of the forward transmitted beam at the exit of the slide stack is imaged, using a single lens in a 2f-2f arrangement, on a charge-coupled-device (CCD) camera.  The CCD camera used here does not need to be of research quality.  Instead, a commercial camera used for ordinary photography works just as well.  In this case, it is important to insure that the CCD readout is electronically accessible, and that the software to do that is available from the camera manufacturer.  The portion of the spatially integrated intensity of backwards scattered light from the slide stack is detected with an optical power meter or an ordinary photodiode detector.  The efficiency with which light is backwards scattered, in relation with the incident beam, is calibrated by replacing the stack of slides with a single high quality mirror coated for the incident optical wavelength and directing the beam back to the power meter.   A nonpolarizing beam splitter cube enables separation of the backscattered intensity from the incident light originating with the helium neon laser.   The 2.54 $\times$ 2.54 $cm^2$ square glass slides are microscope cover slips having a typical thickness on the order of 0.15 $mm$. The cover slips are not optically flat; separate interferometry measurements of light reflected from a single slide showed that they have a significant thickness variation over their surface on the order of the optical wavelength.  Because of these imperfections, a small amount of light is also scattered sideways and away from the edges of the slides.  This light can be seen by eye in a darkened laboratory.  The integrated power of this light was also measured directly with an optical power meter, and was found to be, for a stack of 100 slides, typically less than 10 $\%$  of the optical power incident on the face of the first slide.  Similarly, with the slide stack removed, and no mirror in place, the pixel by pixel and the spatially integrated response of the CCD camera to the incident laser light intensity is calibrated.  The measured and spatially integrated transmitted and reflected optical power is then normalized in such a way that the measurements of these two quantities correspond to the traditional reflection and transmission coefficients in optics.

In a typical experimental run, the calibrating measurements of the transmitted and reflected incident powers $T_0$ and $R_0$ are first made.  A single slide is then placed in the beam, normal to the beam axis, and the reflected total power R, and the transmitted light T is measured.  At the same time, the image of the transmitted light spatial profile is recorded by the CCD camera and saved for later analysis.  A second slide is then added to the beam, and the measurements repeated for N slides, up to N = 100.   In order to have a stable alignment and a common focal plane for the transmitted image, the slides are placed  horizontally in a snugly fitted slide guide, and the stack grows vertically, with the first slide on the bottom of the stack (this defines the focal plane for the images).  In addition, the slides are light, and after placement of an additional slide,  it takes approximately 10 seconds for the slide stack to come to steady state.  This assessment is made by taking a sequence of images with the CCD camera following placement of a slide.  The time scale for reaching steady state did not measurably depend on which slide in the sequence of 100 was placed; for this reason we believe that the relaxation was not due to a significant reconfiguration of the entire stack, but only the placement of a single slide.   Note that we also handled the slides with dust free latex gloves in order to minimize smudging or otherwise contaminating the slide surfaces.

\subsection{Analysis tools}

The mainly theoretical paper by Berry and Klein, \cite{Berry} and also the study by Lu, $\emph{et al.}$ \cite{Lu} (and references found there) are very useful in analysis of one dimensional localization, including analysis of the experimental results obtained in this paper.  Of particular utility is the derivation of the expected law for the normalized transmitted power under conditions of one dimensional localization.  This is  the normalized exponential decay law

\begin{equation}
T(N) = exp(-\xi N)
\label{Eq1}
\end{equation}
where
\begin{equation}
\xi = ln(1/T_1)
\label{Eq2}
\end{equation}
is the inverse localization length as measured by the number of slides. The quantity $T_1$ is the normalized averaged transmission through one slide.  For a single dielectric slide having an index of refraction n, the expression for the single slide transmission, averaged over phases, is

\begin{equation}
T_1 = \frac{16n^2}{(1 + n)^4}.
\label{Eq3}
\end{equation}

A second useful result is the optical equivalent of Ohm's Law, which may be derived in a number of ways, including the transfer matrix approach for intensities rather than amplitudes.  This law applies when there is dominant dephasing in the scattering process such that interferences may be safely ignored.   Ohm's Law in this case is
\begin{equation}
\frac{1}{T} = 1 + N(1 - T_1)/T_1.
\label{Eq4}
\end{equation}
Here T is considered to be the average normalized transmission.  Note the linear dependence on N.

Finally, we point out that first order expansion of the exponential decay law Eq. ($\ref{Eq1}$), along with Eq. ($\ref{Eq2}$), yields the Ohm's Law result Eq. ($\ref{Eq4}$).  This expansion is valid when N is much smaller than the localization length, and when $R_1$ = 1 - $T_1$ is also small, as in the present case.  The physical meaning of this is that the interferences responsible for localization have to build up as the number of glass plates increases up to and beyond the localization length.  We will see in a later section that when this occurs, Eq. ($\ref{Eq4}$) breaks down, indicating transition to the localization regime.

\subsection{Other investigations}
The experimental approach, as described, has also been applied, with similar results, to many different realizations of disorder, and for commercially available glass slides of two different average thickness ranges: 0.13 - 0.17 $mm$ and 0.16 - 0.19 $mm$.  We have also used a number of different convenient light sources, including low-power laser pointers, which provide stable output optical power, and nearly monochromatic stabilized laser diodes.  We also have used the transfer matrix approach to theoretically model many of the results and expected behavior \cite{Berry,Delande} as a function of index of refraction.  We finally point out that there are other related and useful approaches to modeling one dimensional systems.   For instance, vibrational wave packets on a disordered chain of springs and masses may be analyzed in the context of a localization length. \cite{Allen}  For electron transport in one dimension, a disordered lattice model akin to the tight binding model first introduced by Anderson \cite{Anderson1} forms a useful study at the proper level. \cite{Domin}

\section{Results and discussion}
In this section we present our main results and discuss them in the context of light localization in one dimension.
\begin{figure}[th]
\begin{center}
{$\scalebox{0.75}{\includegraphics*{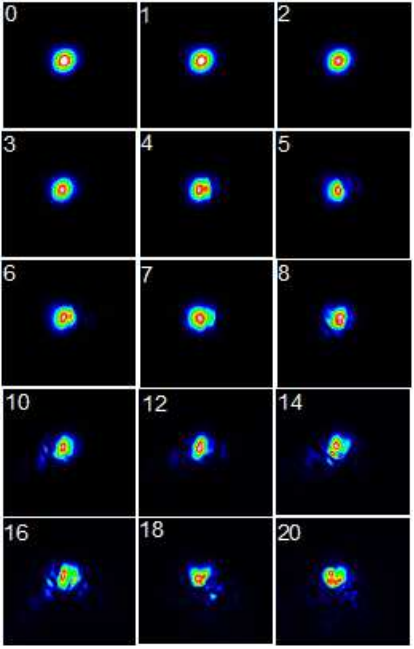}}
$ }
\caption{Characteristic CCD images of the transverse spatial profile of the transmitted light intensity from a stack of a varying number of slides, as indicated on the images. The diameter for these images is on the order of 1 mm. }
\end{center}
\par
\label{fig2}
\end{figure}

We begin with representative transmission images formed by building, as described above, a single stack of 20 slides. The number in the upper left corner of each image is the number of slides in the stack. These images are shown in Figure 2.   One important feature of these transmitted light images is that they maintain their overall qualitative transverse size as the number of slides increases.  This feature is maintained to good approximation up to the maximum number of slides we used; this corresponds to 200 reflecting surfaces in the path of the laser beam.   A second quality is the rather modest development of distortion and speckle structure as the number of slides is increased.  This happens because the thickness of the slides are not transversely uniform.   The fact that we observe these variations means that the phase $nk \delta d$  over the transverse profile of the laser beam can be significant for the slides we use.  Here $\delta d$ is a characteristic thickness variation of a slide and the magnitude of the wave vector $k = 2 \pi / \lambda$.  The real part of the index of refraction of the glass slides is n.  A third important aspect is the development of speckle itself.  This physically means that the scattering and transmission process cannot be strictly one dimensional.  In the case of a true one dimensional system with an incident plane wave beam, we would only observe variations in the overall scale of intensity of each image; the image shape and intensity profile would not change.

\begin{figure}[th]
\begin{center}
{$\scalebox{0.9}{\includegraphics*{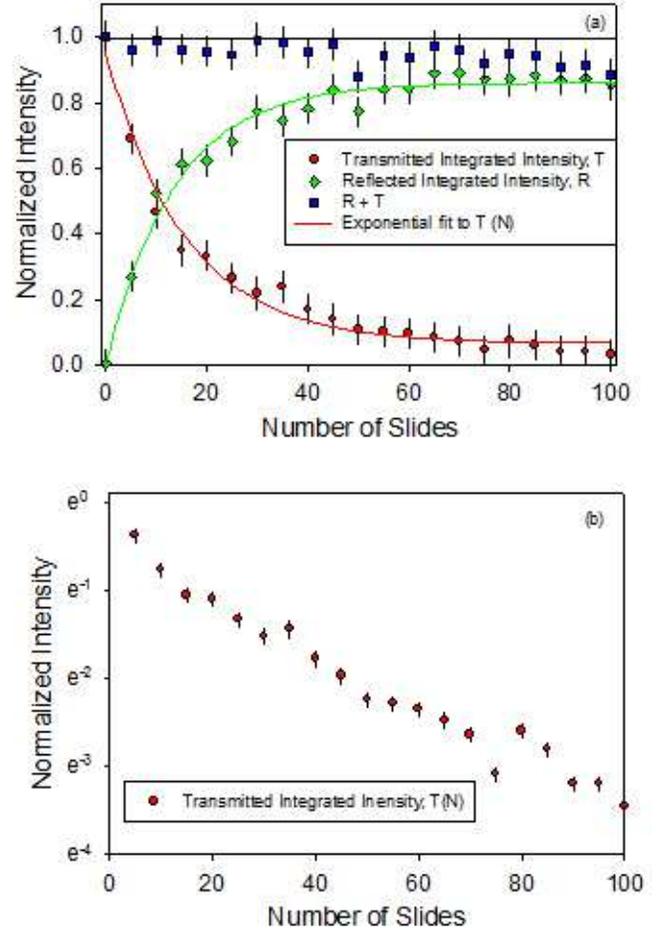}}
$ }
\caption{(a) Variation of the reflection and transmission coefficients with the number of slides N up to N = 100 slides.  The solid curve on the transmission coefficient is an exponential fit to the data.  The sum R + T is also shown, and compared to a reference horizontal curve.  In the absence of losses of light intensity, the quantity R + T would be unity.  The losses in this case are not due to absorption, but instead from light scattered out of the forward and backwards directions. (b) Exponential decay of the transmission coefficient T shown on a natural logarithmic scale. }
\end{center}
\par
\label{fig3}
\end{figure}

The main experimental results of this report are the transmitted and reflected optical powers by the stack of slides, and how these observables evolve as the number of slides is increased.   We show typical such results in Figure 3 for the number of slides N varying from 0 up to 100.   In this figure, the power is normalized in each case to the incident laser beam power, and so we define the transmitted power T as the transmission coefficient, and the reflected power R as a reflection coefficient for the stack as a whole.   With this normalization, the total normalized power T + R would be equal to unity, if no light were lost in the transit through the slide stack.   There are two main possibilities for such light loss.  One of these is optical absorption of light by the slides. However, at the wavelength of light used in these experiments, the absorption coefficient for glass cover slips is normally very small.  However, small amounts of strongly absorbing contaminants on the surfaces could lead to absorption losses nonetheless.   A second possibility is scattering of light out of the viewing area of the optical detectors.   In fact, we observe a small amount of loss, on the order of 10 $\%$ for N = 100 slides.  This loss is due to scattering of light transversely out of the propagation direction. In order to check this, we first observed that there was indeed light scattering sideways out of the slide stack (it could be seen by eye in the darkened lab).   We measured the intensity of this light at several scattering angles for N = 100 and estimated that the spatially integrated power corresponds closely to the loss as seen in Figure 3.  In Figure 3 we then see that the transmission coefficient decreases strongly with increasing N, while the reflection coefficient increases correspondingly with increasing N.  Within the experimental uncertainties, and correcting for transverse scattering losses, the combined transmission and reflection coefficients sum to unity.

The transmission coefficient in Figure 3 decreases exponentially with increasing number of slides N over the full range of the data; the solid curve through T in the figure is an exponential fit to that data, and has a decay constant as a function of N of $\xi$ = 0.064(5) such that T = exp(-$\xi$ N).  The localization length, as measured in terms of the number of slides is then $N_{loc}$ = 14.5.  Similarly, the exponent of the fit to the reflection coefficient R is 0.074(5).  These numbers are consistent with the formulas (see Eqs. 2.1 - 2.3) presented by Berry, $\emph{et al.}$ \cite{Berry}  Within that analysis the determining physical quantity is the average (over phases, or thicknesses) transmission $T_1$ of a single slide.  We have then made measurements of $T_1$ for 100 different slides.  The distribution of those measurements is shown in Figure 4.  There it is seen that measurements of $T_1$ are distributed over the full range of possible values, this being 0.85 - 1.00.  The variations arise from the varying thicknesses of the slides, and the measurements support the conclusion that the variations in slide thickness in the experiments result in good averaging over phases in the measurements of the stacks of slides studied here.

With the data in Figures 2 - 3, we point out an essential physical idea in interpreting these results; localization is in essence a statistical phenomenon.  This subtle point means that as we grow a single realization (stack) of slides we would expect the forward and backwards scattered intensity, as shown in Figure 3, to significantly fluctuate \cite{bert}
as individual slides are added. We further would not be guaranteed a single exponential decay from a single realization.  However, such an exponential decay, and limited noise in the data, is clearly demonstrated in the experimental data.  The key to understanding this result is seen in Figure 2.  The development of speckle in the forward transmitted light means that there are several output spatial modes generated by the transverse imperfections in the slides. Integration or averaging the intensity over these modes smooths the statistical fluctuations in the total intensity, and results in the relatively noise free data in Figure 2.

In Table 1 we summarize the values of $T_1$ determined in different ways. These values are seen to be in very good agreement with each other, even though they are determined by very different methods.  The first column shows the separate measurements as shown in Figure 4, the second and third columns present the results obtained from the fits in Figure 3, while the result labeled $\emph{Theory}$ comes from the averaged expression Eq. 2.4 with an assumed value of the index of refraction n = 1.5.

\begin{table}[h!]
  \centering
   \label{Table1}
   \caption{Average transmission coefficients for a single slide determined by several methods. The result labeled $\emph{Theory}$ comes from the averaged expression Eq. 2.4 with an assumed value of the index of refraction n = 1.5.}
   \begin{tabular}{c c c c}
    \hline
    \hline
      &     & $T_1$  &   \\
      \hline
    Measurement & Transmission   & Reflection & Theory (n=1.5) \\
     \hline
0.91(5)   &   0.938(5)   &   0.929(5)  &   0.922    \\

    \hline
    \hline

  \end{tabular}
 \end{table}

 \begin{figure}[th]
\begin{center}
{$\scalebox{0.9}{\includegraphics*{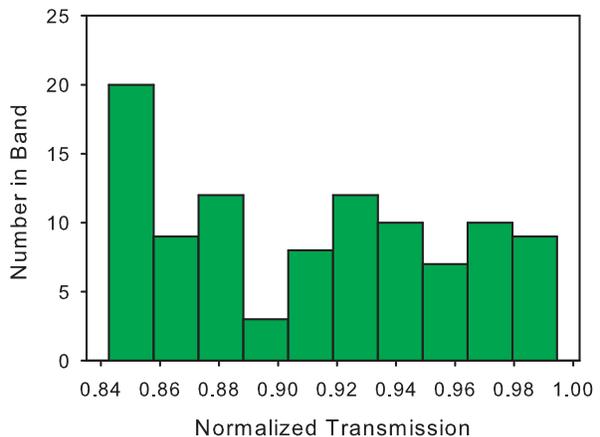}}
$ }
\caption{Histogram of measurements of the normalized transmission $T_1$ for 100 different single slides.   }
\end{center}
\par
\label{fig4}
\end{figure}

The exponential decay seen in Figure 3 is the expected result due to interferences in one dimensional collective scattering  from the stack of slides.  This is one aspect of light localization.   For comparison, if interferences could be ignored, then the multiple scattering of intensities rather than amplitudes would apply.  This is the case when the length scale for randomizing the phase of the scattered light is short enough that localization cannot develop.  For that situation, the transmission varies according to an optical analog of Ohm's Law.  Then the reciprocal transmission $1 / T = 1 + N(1 - T_1)/T_1$; a linear dependence on 1/T with the number of slides N is expected.   Here $T_1$ is the average transmittance of a single slide.  To compare our results to the Ohm's Law prediction, we plot in Figure 5 the data of Figure 3 in this form.  There it can be seen that the transmitted power is indeed not linear with an increasing number of slides.  Of course, as the power decays exponentially with increasing N, there is at first always a linear regime associated with the usual  expansion of the exponential function.  The essential point is that the $\emph{departure}$ from linearity, the failure of an Ohm's Law description, and the corresponding exponential decay with N of the transmission coefficient, all together underscore the fact that inclusion of interferences are essential to properly model the scattering process.

\begin{figure}[th]
\begin{center}
{$\scalebox{0.9}{\includegraphics*{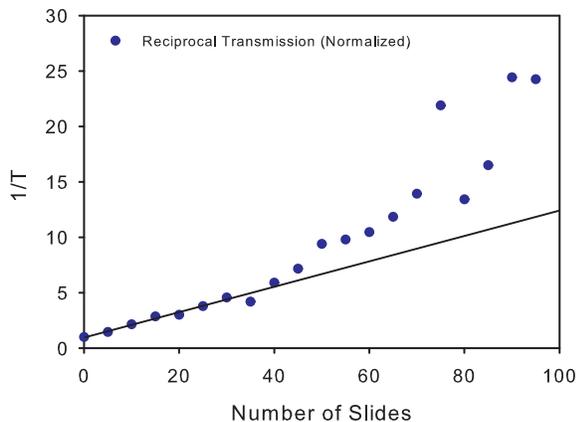}}
$ }
\caption{Graphical illustration of the nonlinearity of the transmitted light power with increasing sample thickness (with N in this case).  Ohm's Law for the transmitted power predicts a linear dependence not seen in the data.  }
\end{center}
\par
\label{fig5}
\end{figure}

Finally, we consider the images of the forward transmission as the number of slides evolves in the range N = 0 - 20.  These are shown in Figure 2.  It is clear from these images, taken in a 2f - 2f optical configuration, that there is evolution of some speckle in the forward transmission.   This means that some of the light scattered by the slides develops a transverse wave vector component that is large enough that the physical system is no longer exactly one dimensional.   Although we did not study this effect in detail, we note that over a large number of experimental runs, the degree of speckle formation varied widely, and yet a single exponent described well both the decay of the transmitted power and the buildup of the backscattered power.  We also did not study development of speckle in the backwards direction, though that would be an interesting avenue for future studies.

\section{Concluding remarks}
This paper is intended as a guide for an upper division undergraduate experimental study of localization of light in a quasi one dimension configuration.  As such we have presented detailed description of an economical and accessible experimental approach to the problem.   We have also provided appropriate references to theoretical papers which are largely on an accessible level for upper level physics or optical engineering majors.  With these tools, we have demonstrated a typical realization of a one dimensional light localization study which may readily be reproduced or elaborated on.   On the experimental side, we have made and presented experimental data on the normalized transmission and reflection coefficients associated with a stack of N slides, and how those quantities evolve as N is increased.   From this data the localization exponent has been determined from the exponential decay of the transmitted power as a function of N.  The decay exponent agrees favorably with one calculated from theoretical expressions and also obtained from separate measurements of the average transmission of a single slide.   We have also shown that an optical version of Ohm's Law, appropriate for incoherent transport, does not describe the experimental data.  Finally, we have shown that as N increases there is a mild development of speckle on the output face of the slide stack.  As speckle would not appear in a purely one dimensional system, this observation may be interpreted as evolution of the experimental configuration towards a quasi one dimensional system.

\section*{Acknowledgements}
We acknowledge financial support of the National Science Foundation (Grant No. NSF-PHY-1068159) and Old Dominion University.

\end{document}